\documentstyle[12pt]{article}
\begin{document}
\title{Causal Interpretation and Quantum Phase Space}
\author{Nuno Costa Dias\footnote{{\it ncdias@mail.telepac.pt}} \\ Jo\~{a}o Nuno Prata\footnote{{\it joao.prata@ulusofona.pt}} \\ {\it Departamento de Matem\'atica} \\
{\it Universidade Lus\'ofona de Humanidades e Tecnologias} \\ {\it Av. Campo Grande, 376, 1749-024 Lisboa, Portugal}}

\maketitle
\begin{abstract}
We show that the de Broglie-Bohm interpretation can be easily implemented in quantum phase space through the method of quasi-distributions. This method establishes a connection with the formalism of the Wigner function. As a by-product, we obtain the rules for evaluating the expectation values and probabilities associated with a general observable in the de Broglie-Bohm formulation. Finally, we discuss some aspects of the dynamics. 
\end{abstract}

\section{Introduction}

Over the years there have been many attempts to grasp the connection between quantum and classical mechanics. One of the main obstacles one immediately faces resides in the fact that the two theories have very disparate formulations. In classical theory the state of the system is completely described by a set of numbers $(\vec x, \vec p)$ at a given time, from which one can predict all future and retrodict all past configurations by solving Hamilton's equations of motion. Morover, it is pressuposed that the system will deterministically pursue these well defined trajectories (solutions of Hamilton's equations) whether a process of measurement is actually performed or not. On the other hand, in quantum theory, and according to standard Copenhagen interpretation, the wavefunction of the system encompasses {\it all} the information about the state of the system that is accessible to us. This description is not deterministic in the sense that the wavefunction only permits probabilistic predictions for the outcome of a measurement. The actual measurement process is problematic, namely in that the system is disturbed by the measurement process resulting in the collapse of the wavefunction. Futhermore, the notion of a well defined trajectory is spoiled and thus the objective existence of matter becomes inextricably linked to the process  of measurement.

Possibly, the most celebrated attempts to formulate classical and quantum mechanics in terms of a common language are the de Broglie-Bohm \cite{Holland1}-\cite{Bohm1} interpretation and the Wigner (or, more generally, the quasi-distribution) formulation \cite{Lee1}-\cite{Flato1} of quantum mechanics. The purpose of this letter is to establish a relation between the two formulations. We will show that the de Broglie-Bohm theory can be implemented through a quasi-distribution in quantum phase space. In doing so, we shall obtain a systematic (albeit mathematically involved) method for computing the local expectation value of a generic operator $\hat A$ in a particular state $|\psi>$. As a by-product, we will write a previously missing formula for evaluating the probabilties associated with a generic observable. Finally, we discuss the Heisenberg picture of the de Broglie-Bohm formulation.

\section{Quasi-distributions in quantum phase space}

Let $\psi (\vec x,t)$ be the wave function representing the state of a one-particle system at any given time $t$. The wave function is a solution of the Schr\"odinger equation:
\begin{equation}
i \hbar \frac{\partial \psi}{\partial t} (\vec x,t)= - \frac{\hbar^2}{2m} \vec{\nabla}_x^2 \psi (\vec x,t) + V(\vec x) \psi(\vec x,t),
\end{equation}
where $V(\vec x)$ is the potential and $\vec x= (x,y,z)$. The momentum $\hat{\vec p}$ is canonically conjugate to $\hat{\vec x}$: $\left[\hat x_i, \hat p_j \right] = i \hbar \delta_{ij}$, $i,j=1,2,3$. According to Cohen's classification \cite{Cohen}, the $f$-quasidistribution function for the pure state $\psi (\vec x,t)$ is given by:
\begin{equation}
F^f (\vec x,\vec p;t) = \frac{1}{(2 \pi)^6} \int d \vec{\xi} \int d \vec{\eta} \int d \vec x' \hspace{0.2 cm} \psi^* \left(\vec x' - \frac{1}{2} \hbar \vec{\eta} ; t \right) \psi \left( \vec x' + \frac{1}{2} \hbar \vec{\eta} ; t \right) f (\vec{\xi}, \vec{\eta} ; t) e^{i \vec{\xi} \cdot ( \vec x' - \vec x)} e^{- i \vec{\eta} \cdot \vec p}.
\end{equation}
The average value of a generic operator $\hat A ( \hat{\vec x}, \hat{\vec p},t)$ can be evaluated according to:
\begin{equation}
< \psi | \hat A (\hat{\vec x}, \hat{\vec p},t ) | \psi > = \int d \vec x \int d \vec p \hspace{0.2 cm} \tilde A_f (\vec x,\vec p,t) F^f (\vec x,\vec p,t),
\end{equation}
where $\tilde A_f (\vec x,\vec p,t)$ is a c-function known as the "$f$-symbol" associated with the operator $\hat A$:
\begin{equation}
V_f \left( \hat A (\hat{\vec x}, \hat{\vec p} ) \right) = \tilde A_f (\vec x,\vec p,t) \equiv \left(\frac{\hbar}{2 \pi}\right)^3 \int d \vec{\xi} \int d \vec{\eta} \hspace{0.2 cm} Tr \left\{ \hat A (\hat{\vec x}, \hat{\vec p}) e^{i \vec{\xi} \cdot \hat{\vec x} + i \vec{\eta} \cdot  \hat{\vec p}} f^{-1} ( \vec{\xi}, \vec{\eta} ,t)  \right\} e^{-i \vec{\xi} \cdot \vec x -i \vec{\eta} \cdot \vec p}.
\end{equation}
The application $V_f : \hat{\cal A} ({\cal H}) \to {\cal A} (T^* M)$ is called the "{\it $f$-map}" \cite{Lee1} and, at each time $t$, can be regarded as an isomorphism between the quantum algebra of operators, $\hat{\cal A} ({\cal H})$, acting on the Hilbert space ${\cal H}$, and a "classical" algebra of c-functions, ${\cal A} (T^*M)$ in the phase space $T^* M$. Consequently, the map $V_f$ allows for a formulation of quantum mechanics in terms of "classical-like" objects.

\noindent
The most celebrated example corresponds to $f_W (\vec{\xi}, \vec{\eta},t) =1$, in which case (4) yields the Weyl-symbol \cite{Weyl}:
\begin{equation}
V_W \left( \hat A (\hat{\vec x}, \hat{\vec p} ) \right) = \tilde A_W (\vec x,\vec p) \equiv \left(\frac{\hbar}{2 \pi}\right)^3 \int d \vec{\xi} \int d \vec{\eta} \hspace{0.2 cm} Tr \left\{ \hat A (\hat{\vec x}, \hat{\vec p}) e^{i \vec{\xi} \cdot \hat{\vec x} + i \vec{\eta} \cdot \hat{\vec p}} \right\} e^{-i \vec{\xi} \cdot \vec x -i \vec{\eta} \cdot \vec p},
\end{equation}
where $V_W$ is the Weyl map. The corresponding quasi-distribution function (2) is proportional to the Weyl-symbol of the quantum density matrix $\hat{\rho} = | \psi >< \psi|$. It is named the Wigner function \cite{Wigner}:   
\begin{equation}
F^W (\vec x,\vec p,t) \equiv \frac{1}{(2 \pi \hbar)^3} V_W ( \hat \rho) = \frac{1}{(2 \pi)^3} \int d \vec y e^{-i \vec y \cdot \vec p} \psi^* \left( \vec x- \frac{1}{2} \hbar \vec y, t \right) \psi \left( \vec x + \frac{1}{2} \hbar \vec y, t \right).
\end{equation}
It is easy to check that it is a real function. The Weyl-symbol of the operator product and of the quantum commutator are known as the $*$-product and the Moyal bracket \cite{Moyal}, respectively:
\begin{equation}
\begin{array}{l l l}
V_W \left( \hat A \cdot \hat B \right)  & = \tilde A_W *_W \tilde B_W & =\tilde A_W (\vec x,\vec p) e^{\frac{i \hbar}{2} {\hat{\cal J}}} \tilde B_W (\vec x,\vec p) ,\\
& & \\ 
V_W \left( \left[ \hat A , \hat B \right] \right)  & = \left[ \tilde A_W , \tilde B_W \right]_M & = 2 i \tilde A_W (\vec x, \vec p) \sin \left(\frac{\hbar}{2} {\hat{\cal J}} \right) \tilde B_W (\vec x, \vec p),
\end{array}
\end{equation}
where ${\hat{\cal J}}$ is the "{\it Poisson}" operator: $
{\hat{\cal J}} \equiv \sum_{i=1}^3 \left( \frac{ {\buildrel { \leftarrow}\over\partial}}{\partial q_i} \frac{ {\buildrel { 
\rightarrow}\over\partial}}{\partial p_i} -  \frac{{\buildrel { \leftarrow}\over\partial}}{\partial p_i}  \frac{{\buildrel { 
\rightarrow}\over\partial}}{\partial q_i} \right)  $, 
the derivatives ${\buildrel { \leftarrow}\over\partial}$ and ${\buildrel { \rightarrow}\over\partial}$ acting on $\tilde A_W$ and $\tilde B_W$, respectively. From eq.(1) it is possible to obtain the differential equation governing the dynamics of $F^W$:
\begin{equation}
\frac{\partial F^W}{\partial t} = \frac{1}{i \hbar} \left[\tilde H_W, F^W \right]_M , \qquad \tilde H_W(\vec x,\vec p) = \frac{{\vec p}^{\hspace{0.1cm} 2}}{2m} + V(\vec x),
\end{equation}
where $\tilde H_W(\vec x,\vec p)$ is the Weyl symbol of the quantum Hamiltonian. Eq.(8) can be regarded as a deformation of the classical Liouville equation. A remarkable theorem by Baker \cite{Baker} states that provided a real function $F^W$ satisfies (8) and some normalization requirement, then there exists a complex function $\psi$ which (i) is a solution of the Schr\"odinger equation (1) and (ii) is related to $F^W$ according to eq.(6). Consequently, the Wigner function encompasses the same information about the state of the system as does the wave function. Indeed, it is possible to evaluate all the mean values (using eq.(3)) and also all probabilities according to the following procedure. 

Consider a complete set of commuting observables $\hat A_1, \cdots , \hat A_n$, and let $|a_1, \cdots, a_n>$ be a general eigenstate, satisfying: $\hat A_i |a_1, \cdots, a_n> = a_i |a_1, \cdots, a_n>$, $(i=1, \cdots, n)$. The projector $|a_1, \cdots, a_n><a_1, \cdots, a_n|$ also satisfies the previous eigenvalue equation:
\begin{equation}
\hat A_i |a_1, \cdots, a_n><a_1, \cdots, a_n| = a_i |a_1, \cdots, a_n><a_1, \cdots, a_n|, \hspace{0.5 cm} i=1, \cdots,n .
\end{equation}
The probability that a measurement of the operator $\hat A_i$ yield the eigenvalue $a_i$ is given by\footnote{If the spectrum is discrete, the integrals are replaced by sums.}:
\begin{equation}
\int da_1 \cdots \int d a_{i-1} \int d a_{i+1} \cdots \int da_n \hspace{0.2 cm} | <\psi | a_1, \cdots , a_n>|^2  = Tr \left\{ \hat{\rho} |a_i><a_i| \right\},
\end{equation}
where $|a_i><a_i|$ is also a projector, which reads:
\begin{equation}
|a_i><a_i| \equiv \int da_1 \cdots \int d a_{i-1} \int d a_{i+1} \cdots \int da_n \hspace{0.2 cm} |a_1, \cdots, a_n><a_1, \cdots, a_n|.
\end{equation}
Notice that $|a_i>< a_i|$ is also a solution of the eigenvalue equation (9). 

In the formalism of the Wigner function, one introduces the so called stargenfunctions \cite{Fairlie1}-\cite{Flato1}:
\begin{equation}
\begin{array}{l}
\delta_*^W \left( \tilde A_{1W} - a_1 \right) *_W \cdots *_W \delta_*^W \left( \tilde A_{nW} -a_n \right) = V_W \left( |a_1, \cdots, a_n><a_1, \cdots, a_n| \right)\\
\\
\delta_*^W \left( \tilde A_{iW} - a_i \right) = V_W \left( |a_i><a_i| \right) = \frac{1}{2 \pi} \int_{- \infty}^{+ \infty}  dk  \hspace{0.2 cm} e_{*_W}^{i  k (\tilde A_{iW} -a_i)},
\end{array}
\end{equation}
where $\tilde A_{iW} = V_W \left( \hat A_i \right)$ $(i=1, \cdots , n)$. The $*$-exponential is defined by $e_{*_W}^{A (\vec x, \vec p)} =  \sum_{n=0}^{\infty} \frac{1}{n!} \left( A (\vec x, \vec p) \right)^{n*_W}$, where $A^{n*_W}=A*_W A^{n-1*_W}$ and $A^{0*_W}=1$. These stargenfunctions are solutions of the stargenvalue equations:
\begin{equation}
\begin{array}{l}
\tilde A_{iW} *_W \left\{ \delta_*^W \left( \tilde A_{1W} - a_1 \right) *_W \cdots *_W \delta_*^W \left( \tilde A_{nW} -a_n \right) \right\} = a_i \delta_*^W \left( \tilde A_{1W} - a_1 \right) *_W \cdots *_W \delta_*^W \left( \tilde A_{nW} -a_n \right)\\
\\
\tilde A_{iW} *_W \delta_*^W \left( \tilde A_{iW} - a_i \right)= a_i \delta_*^W \left( \tilde A_{iW} - a_i \right),
\end{array}
\end{equation}
which are the Wigner-Weyl counterparts of the eigenvalue equations (9). Moreover, the probability (10) can be evaluated according to the following suggestive formula\footnote{Here $A_i$ denotes the observable $\hat A_i$ irrespective of the particular representation chosen, i.e. irrespective of the particular function $f \left( \vec{\xi}, \vec{\eta},t \right)$.}:
\begin{equation}
{\cal P}_1 \left( A_i (t) =  a_i \right) =  \int d \vec x \int d \vec p \hspace{0.2 cm} F^W (\vec x,\vec p ,t) \delta^W_* (\tilde A_{iW}-a_i).
\end{equation}
In particular, for $\hat A = \hat{\vec x}, \hat{\vec p}$, it can be shown that: $\delta_*^W (x_i - x_i') = \delta (x_i - x_i')$ and $\delta_*^W (p_i - p_i') = \delta (p_i - p_i')$ $(i=1,2,3)$, and we thus get the classical-like results:
\begin{equation}
\begin{array}{l l}
{\cal P}_1 (\vec x= \vec x') & = \int d \vec x \int d \vec p \hspace{0.2 cm} F^W (\vec x,\vec p) \delta^3( \vec x- \vec x') =  |\psi (\vec x)|^2,\\
& \\
{\cal P}_1 (\vec p= \vec p') & = \int d \vec x \int d \vec p \hspace{0.2 cm} F^W ( \vec x,\vec p) \delta^3( \vec p- \vec p') =  |\phi (\vec p)|^2,
\end{array}
\end{equation}
where $\phi (\vec p)$ is the Fourier transform of $\psi(\vec x)$. Notice the remarkable similarities between the function $\delta^W_*$ and the ordinary $\delta$-function. Not only does it satisfy eqs.(13,14) and:
\begin{equation}
\frac{\int d \vec x \int d \vec p \hspace{0.2 cm} \tilde A_W( \vec x , \vec p)  \delta^W_* (\tilde A_W-a)}{\int d \vec x \int d \vec p \hspace{0.2 cm}\delta^W_* (\tilde A_W-a)}  = a,
\end{equation}
it is also an $\hbar$-deformation of the $\delta$-function: $\delta^W_* (\tilde A_W-a) = \delta (\tilde A_W -a) + {\cal O} (\hbar)$. It therefore works as a $\delta$-function in quantum *-phase space. For these reasons it is called the *-delta function.  

From eqs.(3,15) one could be tempted to interpret the Wigner function as a probability distribution in phase space. However, this interpretation is immediately spoiled, if one realizes that it can take on negative values. It is a well known fact, that some undesired property will hinder a full classical interpretation for any quasi-distribution. This is usually interpreted as a manifestation of Heisenberg's uncertainty principle.

Finally, the Wigner-Weyl formalism can be regarded as a basis for other quasi-distributions (i.e. with $f(\vec{\xi} , \vec{\eta},t) \ne 1$). It can be checked (cf.(2,4)) that the $f$-symbol of a certain observable and the corresponding quasi-distribution can be obtained from the Weyl symbol and the Wigner function, according to the followig prescription \cite{Lee1, Cohen}:
\begin{equation}
\left\{
\begin{array}{l}
\tilde A_f ( \vec x, \vec p, t) = f^{-1} \left( - i \vec{\nabla}_x, - i \vec{\nabla}_p ,t \right) \tilde A_W (\vec x, \vec p),\\
\\
F^f ( \vec x, \vec p, t) = f \left(  i \vec{\nabla}_x,  i \vec{\nabla}_p ,t \right) F^W (\vec x, \vec p,t),
\end{array}
\right.
\end{equation}
where $\vec{\nabla}_p = \left( \frac{\partial}{\partial p_x} , \frac{\partial}{\partial p_y}, \frac{\partial}{\partial p_z} \right)$.
This permits us to write the counterpart of eqs.(7), (12), (13) and (14) for different choices of function $f( \vec{\xi}, \vec{\eta},t)$. Starting with equation (7), we have:
\begin{equation}
\begin{array}{c}
\tilde A_f *_f \tilde B_f = f^{-1} \left( - i \vec{\nabla}_x, - i \vec{\nabla}_p,t \right) \tilde A_W *_W \tilde B_W =  f^{-1} \left( - i \vec{\nabla}^A_x - i \vec{\nabla}^B_x,  - i \vec{\nabla}^A_p - i \vec{\nabla}^B_p ,t \right) \times\\
\\
\times  f \left( - i \vec{\nabla}^A_x ,  - i \vec{\nabla}^A_p ,t \right) f \left( - i \vec{\nabla}^B_x ,  - i \vec{\nabla}^B_p ,t \right) \exp \left[\frac{i \hbar}{2} \left( \vec{\nabla}^A_x \cdot  \vec{\nabla}^B_p - \vec{\nabla}^A_p \cdot  \vec{\nabla}^B_x \right) \right] \tilde A_f \tilde B_f,
\end{array}
\end{equation}
where the superscripts $A$ and $B$ mean that the derivatives act on $\tilde A_f$ and $\tilde B_f$ respectively. Using this definition of the star-product $*_f$, we can further define the $f$-stargenfunction:
\begin{equation}
\delta^f_* (\tilde A_{f} -a)= \frac{1}{2 \pi} \int_{- \infty}^{+ \infty} d k \hspace{0.2 cm} e_{*_f}^{ik (\tilde A_{f} -a) },
\end{equation}
which can easily be shown to be the $f$-symbol of $|a><a|$ (cf.(4,11,12)). Furthermore, from (13), (14) and (17) it follows that: 
\begin{equation}
\begin{array}{l}
\tilde A_{f}  *_f \delta^f_* (\tilde A_{f} -a) = a \delta^f_* (\tilde A_{f} -a)\\
\\
{\cal P}_1 \left( A (t) =a \right) = \int d \vec x \int d \vec p \hspace{0.2 cm} F^f (\vec x, \vec p ,t) \delta^f_* \left(\tilde A_{f} (t) - a \right).
\end{array}
\end{equation}
  
\section{Causal interpretation}

In the de Broglie-Bohm interpretation \cite{Holland1}-\cite{Bohm1} of quantum mechanics, the wave function is written in the form:
\begin{equation}
\psi (\vec x,t) = R(\vec x,t) \exp \left( \frac{i}{ \hbar} S(\vec x,t) \right),
\end{equation}
where $R(\vec x,t)$ and $S(\vec x,t)$ are some real functions. Substituting this expression in eq.(1) we obtain the dynamics of $R$ and $S$:
\begin{equation}
\left\{
\begin{array}{l}
\frac{\partial R}{\partial t} = -\frac{1}{2m} \left[ R \vec{\nabla}_x^2 S + 2 \vec{\nabla}_x R \cdot \vec{\nabla}_x S \right],\\
\\
\frac{\partial S}{\partial t} = - \frac{1}{2m} \left(\vec{\nabla}_x S \right)^2 - V(\vec x) + \frac{\hbar^2}{2m} \frac{1}{R} \vec{\nabla}_x^2 R.
\end{array}
\right.
\end{equation}
The distribution ${\cal P}_2 (\vec x) \equiv |\psi (\vec x)|^2 = R^2 (\vec x)$ corresponds to the probability of finding the particle at $\vec x$. In terms of ${\cal P}_2$, we can rewrite eq.(22) as:
\begin{equation}
\left\{
\begin{array}{l}
\frac{\partial {\cal P}_2}{\partial t} +  \vec{\nabla}_x \cdot \left(\frac{{\cal P}_2}{m} \vec{\nabla}_x S \right) =0 ,\\
\\
\frac{\partial S}{\partial t} + \frac{1}{2m} \left(\vec{\nabla}_x S \right)^2 + V(\vec x) - \frac{\hbar^2}{4m} \left[ \frac{1}{{\cal P}_2} \vec{\nabla}_x^2 {\cal P}_2 - \frac{1}{2} \frac{1}{{\cal P}_2^2} \left( \vec{\nabla}_x {\cal P}_2 \right)^2 \right] =0.
\end{array}
\right.
\end{equation}
The first equation is a statement of probability conservation, with associated flux $\frac{{\cal P}_2}{m} \vec{\nabla}_x S $. The second equation is interpreted as a Hamilton-Jacobi equation. The solution $S(\vec x,t)$ corresponds to a set of trajectories, known as Bohmian trajectories, stemming from some original position $\vec x$ with probability ${\cal P}_2 (\vec x) $, for a particle under the influence of a classical potential $V(\vec x)$ and a quantum potential:
\begin{equation}
Q(\vec x,t) \equiv - \frac{\hbar^2}{2m} \frac{1}{R} \vec{\nabla}_x^2 R = - \frac{\hbar^2}{4m} \left[ \frac{1}{{\cal P}_2} \vec{\nabla}_x^2 {\cal P}_2 - \frac{1}{2} \frac{1}{{\cal P}_2^2} \left( \vec{\nabla}_x {\cal P}_2\right)^2 \right].
\end{equation}
Morover, the momentum $\vec p$ is subject to the constraint:
\begin{equation}
\vec p = \vec{\nabla}_x S,
\end{equation}
holding at all times (including $t=0$). We can thus conclude that, for a pure state, the quantum phase space distribution is given by \cite{Holland1}:
\begin{equation}
F^B (\vec x,\vec p;t) \equiv {\cal P}_2 (\vec x,t) \delta^3 \left( \vec p - \vec{\nabla}_x S (\vec x,t) \right).
\end{equation}
The dynamics of $F^B$ is obtained from eqs.(22,23):
\begin{equation}
\frac{\partial F^B}{\partial t} (\vec x,\vec p;t) = \left\{ H(\vec x,\vec p) + Q(\vec x;t) , F^B (\vec x,\vec p ;t) \right\}_P,
\end{equation}
where $H$ is given by eq.(8) and $\left\{ , \right\}_P$ is the Poisson bracket. $F^B$ is positive defined, real and has the marginal distribution:
\begin{equation}
{\cal P}_2 (\vec x;t) = R^2 (\vec x;t) = \int d \vec p \hspace{0.2 cm} F^B (\vec x,\vec p;t),
\end{equation}
which coincides with ${\cal P}_1 (\vec x;t)$ (eq.(15)). Integration over the positions yields the momentum probability distribution ${\cal P}_2 (\vec p,t)$. However, this distribution is different from the one in eq.(15). In the de Broglie-Bohm formulation it corresponds to the true momentum of the particle \cite{Holland1}. In order to distinguish the two, we use the subscripts 1 and 2.  

Mean values of observables are evaluated according to the formula:
\begin{equation}
<\hat A> = \int d \vec x \int d \vec p \hspace{0.2 cm} A_B (\vec x,t) F^B (\vec x, \vec p,t),
\end{equation}
where the "local expectation value" $A_B$ is given by \cite{Holland1}:
\begin{equation}
A_B (\vec x, t) = {\cal R}e \left( \frac{\psi^* (\vec x,t) (\hat A \psi ) (\vec x,t)}{| \psi (\vec x,t)|^2} \right).
\end{equation}
With this definition, we get for instance:
\begin{equation}
\left\{
\begin{array}{l l}
\vec p_B (\vec x,t) = \vec{\nabla}_x S & \mbox{(Momentum)} \\
& \\
H_B (\vec x, t) = \frac{(\vec{\nabla}_x S)^2 }{2m}  + Q (\vec x, t)+ V(\vec x) & \mbox{(Energy)}\\
& \\
\vec L_B (\vec x, t) = \vec x \times \vec{\nabla}_x S & \mbox{(Angular momentum)}\\
& \\
\vec L_B^2 (\vec x ,t) = \left( \vec x \times \vec{\nabla}_x S \right)^2 - \frac{\hbar^2}{R} \left( \vec x \times \vec{\nabla}_x \right)^2 R &  
\end{array}
\right.
\end{equation}

It is important to stress that these objects are not $f$-symbols of the corresponding quantum mechanical operators. In fact, and contrary to ordinary $f$-symbols, these objects live in configuration space. A second remark, this time regarding the dynamics, is in order. The classical looking equation (27) does not yield the complete time evolution. This is because if we want $F^B$ to preserve its form with time (and this is crucial for the causal interpretation), we have to evolve the local expectation values as well. We shall come back to this later.

\section{De Broglie-Bohm formulation as a quasi-distribution}

\subsection{Quasi-distribution and Bohm-symbols}

In this section we show that eq.(26) can be obtained from eq.(2) for a particular choice of the function $f$:
\begin{equation}
f_B ( \vec{\xi}, \vec{\eta} ; t) = \frac{\int d \vec u | \psi (\vec u ;t)|^2 \exp \left(i \vec{\eta} \cdot  \vec{\nabla}_u S (\vec u;t) + i \vec{\xi} \cdot \vec u \right) }{\int d \vec v \psi^* \left(\vec v- \frac{\hbar}{2} \vec{\eta };t \right) \psi \left(\vec v + \frac{ \hbar}{2} \vec{\eta } ; t \right) e^{i \vec{\xi} \cdot \vec v}}.
\end{equation}
This is the most important formula of this work. Substituting in eq.(2), we obtain:
\begin{equation}
\begin{array}{c}
F^B (\vec x,\vec p;t) = \frac{1}{(2 \pi)^6} \int d \vec{\xi} \int d \vec{\eta} \int d \vec u \hspace{0.2 cm} e^{- i \vec{\xi} \cdot \vec  x - i \vec{\eta} \cdot \vec  p} | \psi (\vec u ,t)|^2 e^{i \vec{\eta} \cdot \vec{\nabla}_u S (\vec u ,t) + i \vec{\xi} \cdot \vec u} = \\
\\
= \frac{1}{(2 \pi)^3} \int d \vec{\eta} \int d \vec u \hspace{0.2 cm} e^{- i \vec{\eta} \cdot \vec p} | \psi (\vec u ,t)|^2 e^{i  \vec{\eta} \cdot \vec{\nabla}_u S (\vec u ,t)} \delta^3 (\vec u- \vec x) = | \psi (\vec x ,t) |^2 \delta^3 \left( \vec p - \vec{\nabla}_x S (\vec x,t) \right),
\end{array}
\end{equation}
and we do indeed recover eq.(26). Moreover the Bohm-symbol of an operator $\hat A ( \hat{\vec x}, \hat{\vec p})$ can be obtained from the Weyl-symbol according to eq.(17):
\begin{equation}
\tilde A_B (\vec x, \vec p,t) = V_B \left( \hat A (\hat{\vec x} , \hat{\vec p}) \right) = f_B^{-1} \left( - i \vec{\nabla}_x, - i \vec{\nabla}_p,t \right) \tilde A_W (\vec x, \vec p).
\end{equation}
If we expand eq.(32) in powers of $\vec{\eta}$, we get:
\begin{equation}
f^{-1}_B (\vec{\xi}, \vec{\eta} ;t ) = 1 + \frac{\hbar^2}{4 } <e^{i \vec{\xi} \cdot \vec x}>^{-1} < \left(\partial_i \partial_j \ln R \right) e^{i \vec{\xi} \cdot \vec x} > \eta_i \eta_j + \cdots
\end{equation} 
where we used the notation: $<g ( \vec x)> = \int R^2 ( \vec x) g( \vec x) d \vec x$  for any function $g(\vec x)$.
From the previous equation and eq.(34), we conclude that if $\hat A$ is independent of $\hat{\vec p}$ or linear in $\hat{\vec p}$, then $\tilde A_B = \tilde A_W$. Consequently:
\begin{equation}
\left\{
\begin{array}{l l}
V_B \left(V( \hat{\vec x}) \right) & =  V(\vec x)\\
& \\
V_B \left(\hat p_i \right) & = p_i\\
& \\
V_B \left(\hat L_i \right) & = \sum_{j,k} \epsilon_{ijk} x_j * p_k = \sum_{j,k} \epsilon_{ijk}  \left( x_j  p_k + \frac{i \hbar}{2} \delta_{jk} \right) = (\vec x \times \vec p)_i
\end{array}
\right.
\end{equation}
Let us go back to the remark made at the end of the previous section. The Bohm-symbols are phase space quantities and must therefore differ from the local expectation values in equations (30,31), which live in configuration space. This notwithstanding both sets will yield identical predictions, as we shall see.

For operators $\hat B ( \hat x, \hat p)$ quadratic in the momentum, the Bohm-symbol becomes more involved. As a concrete example, consider the kinetic energy, $\hat T (\hat{\vec p}) = \frac{\hat{\vec p}^{\hspace{0.1cm}2}}{2m}$. Since $\hat T$ is independent of $\hat{\vec x}$, we obtain from (34,35):
\begin{equation}
\begin{array}{c}
V_B \left(\hat T \right) = \frac{\vec p^{\hspace{0.1cm}2}}{2m} -\frac{\hbar^2}{4} < \left( \partial_i \partial_j \ln R \right) > \frac{\partial}{\partial p_i} \frac{\partial}{\partial p_j} \frac{\vec p^2}{2m}
= \frac{\vec p^{\hspace{0.1cm}2}}{2m} - \frac{\hbar^2}{4m} < \frac{\vec{\nabla}_x^2 R}{R} - \frac{1}{R^2} (\vec{\nabla}_x R)^2 > =\\
\\
= \frac{\vec p^{\hspace{0.1cm}2}}{2m} - \frac{\hbar^2}{2m} < \frac{\vec{\nabla}_x^2 R}{R}> = \frac{\vec p^2}{2m} + <Q(\vec x;t)>.
\end{array}
\end{equation}
The mean total energy is then:
\begin{equation}
\begin{array}{c}
< \hat H (\hat{\vec x}, \hat{\vec p})> = \int d \vec x \int d \vec p \left\{ \frac{\vec p^{\hspace{0.1cm}2}}{2m} - \int d \vec u \hspace{0.2 cm} R^2 (\vec u ,t) Q (\vec u,t) + V(\vec x) \right\} R^2 (\vec x,t)  \delta^3 \left(\vec p - \vec{\nabla}_x S (\vec x,t) \right) =\\ 
\\
= \int d \vec x R^2 (\vec x ,t) \left\{ \frac{ \left( \vec{\nabla}_x S (\vec x,t) \right)^2}{2m} + V(\vec x) + Q (\vec x,t) \right\} ,
\end{array}
\end{equation}
as expected. Now let us consider $\hat{\vec L}^2$. Since $V_W ( \hat{\vec L}^2) = (\vec x \times \vec p)^2 - \frac{3}{2} \hbar^2$, we get\footnote{Sum over repeated indices is understood and $\partial_i = \frac{\partial}{\partial x_i}$.}:
$$
\begin{array}{c}
V_B ( \hat{\vec L}^2 ) = (\vec x \times \vec p)^2 - \frac{3}{2} \hbar^2 - \frac{\hbar^2}{4} < e^{\vec x \cdot \vec{\nabla}_x}>^{-1} < \left(\partial_i \partial_j \ln R \right) e^{ \vec x \cdot \vec{\nabla}_x} >  \frac{\partial}{\partial p_i} \frac{\partial}{\partial p_j} (\vec x \times \vec p)^2=\\
\\
= (\vec x \times \vec p)^2 - \frac{3}{2} \hbar^2 - \frac{\hbar^2}{2} \left\{ <\partial_i \partial_j \ln R > + \left[ <x_k \partial_i \partial_j \ln R > - <x_k>  <\partial_i \partial_j \ln R > \right] \frac{\partial}{\partial x_k} + \right.\\
\\
 + \frac{1}{2} \left[  <x_k x_l \partial_i \partial_j \ln R > -2 <x_k>  <x_l \partial_i \partial_j \ln R >  + \right.\\
\\
\left. \left. + 2 <x_k> <x_l>  <\partial_i \partial_j \ln R > - <x_k x_l>  <\partial_i \partial_j \ln R > \right] \frac{\partial}{\partial x_k} \frac{\partial}{\partial x_l} \right\} 
\left(\delta_{ij} \vec x^{ \hspace{0.1 cm} 2} - x_i x_j \right) =\\
\\
= (\vec x \times \vec p)^2 - \frac{3}{2} \hbar^2 + \hbar^2 < \vec{\nabla}^2 \ln R > \left[ - \frac{1}{2} \left( \vec x^{ \hspace{0.1 cm} 2} - < \vec x^{ \hspace{0.1 cm} 2} > \right) + \left( x_i - <x_i> \right)  <x_i> \right]+\\
\\
+ \hbar^2 <\partial_i \partial_j \ln R > \left[ \frac{1}{2} \left( x_i x_j  - <x_i x_j  > \right) - \left( x_j - <x_j> \right)  <x_i> \right] \\
\\
- \hbar^2 \left( < x_i \vec{\nabla}^2 \ln R > - <x_j \partial_i \partial_j \ln R > \right) \times  \left( x_i - <x_i> \right) + \frac{\hbar^2}{2} <x_i x_j \partial_i \partial_j \ln R > -  \frac{\hbar^2}{2} <\vec x^{\hspace{0.1 cm} 2} \vec{\nabla}^2 \ln R > .
\end{array}
$$
The previous expression is the Bohm symbol associated with $\hat{\vec L}^2$, which is strikingly different from its local expectation value (31). We now show that the two predictions for the mean value of $\hat{\vec L}^2$ coincide. 
$$
\begin{array}{c}
<\hat{\vec L}^2 > = \int d \vec x \int d \vec p \hspace{0.2 cm} V_B ( \hat{\vec L}^2 ) R^2 (\vec  x) \delta^3 \left( \vec p - \vec{\nabla}_x S \right) = < (\vec x \times \vec{\nabla_x} S )^2 > - \frac{3}{2} \hbar^2 +  \frac{\hbar^2}{2} <x_i x_j \partial_i \partial_j \ln R > \\
\\
-  \frac{\hbar^2}{2} <\vec x{ \hspace{0.1 cm} 2} \vec{\nabla}^2 \ln R >= \int d \vec x \hspace{0.2 cm} R^2 (\vec x) \left\{(\vec x \times \vec{\nabla_x} S )^2  - \frac{3}{2} \hbar^2 +  \frac{\hbar^2}{2} x_i x_j \partial_i \partial_j \ln R  -  \frac{\hbar^2}{2} \vec x^{ \hspace{0.1 cm} 2} \vec{\nabla}^2 \ln R \right\}.
\end{array}
$$
Taking into account that $\frac{3}{2} \hbar^2 = \frac{\hbar^2}{2} \vec{\nabla}_x \cdot \vec x$, we get after a few integrations by parts:
\begin{equation}  
<\hat{\vec L}^2 > =\int d \vec x \hspace{0.2 cm} R^2 (\vec x) \left\{(\vec x \times \vec{\nabla_x} S )^2 + 2 \hbar^2 \frac{1}{R} ( \vec x \cdot \vec{\nabla}_x) R + \hbar^2 x_i x_j  \frac{1}{R} \frac{\partial^2 R}{\partial x_i \partial x_j} - \hbar^2  \vec x^{ \hspace{0.1 cm} 2} \frac{1}{R} \vec{\nabla}_x^2 R \right\}.
\end{equation}
The latter three terms can be rewritten as $- \hbar^2 \frac{1}{R} \left( \vec x \times \vec{\nabla}_x \right)^2 R$, and we thus recover eq.(31).

\subsection{Stargenfunctions and probabilities}

In order to compute of the probability functionals associated with projectors of the form $|a><a|$, we have to determine the corresponding $*$-genfunctions. For the position, the eigenvalue equation is: $\hat{\vec x} |\psi>< \psi| = \vec x' |\psi>< \psi| \Longrightarrow |\psi>< \psi| = |\vec x'><\vec x'|$. If we apply the Bohm-map (34) to the previous equation, we obtain: $\vec x *_B G^B (\vec x, \vec p)  = \vec x' G^B (\vec x, \vec p)$, where the starproduct $*_B$ is given by eq.(18) with $f=f_B$. The solution is:
\begin{equation}
G^B (\vec x, \vec p) = V_B \left( | \vec x'>< \vec x'| \right) =\delta^B_* (\vec x - \vec x') =\delta^3 (\vec x - \vec x').
\end{equation}
The $*$-genfunction is thus $\delta^3 (\vec x - \vec x')$ and consequently:
\begin{equation}
{\cal P}_1 (\vec x',t) = \int d \vec x \int d \vec p \hspace{0.2 cm} F^B (\vec x, \vec p,t) \delta^3 (\vec x - \vec x') = R^2  (\vec x',t).
\end{equation}
This means that ${\cal P}_1 (\vec x,t)$ is indeed equal to ${\cal P}_2 (\vec x,t)$ (28). However, the $*$-genvalue equation for $\hat{\vec p}$ is more involved: $\hat{\vec p} | \phi>< \phi| = \vec p' | \phi>< \phi| \Longrightarrow | \phi>< \phi| = |\vec p'>< \vec p'|$. Upon application of the Bohm-map, we obtain: $\vec p *_B \Theta^B (\vec x, \vec p ,t) =\vec  p' \Theta^B (\vec x, \vec p ,t)$, with solution:
\begin{equation}
\Theta^B (\vec x, \vec p,t)= V_B \left( | \vec p'>< \vec p'| \right) =\delta_*^B (\vec p - \vec p',t) = f_B^{-1} \left( 0 , - i \vec{\nabla}_p,t \right) \delta^3 (\vec p - \vec p'),
\end{equation}
which is not identical to $\delta^3 (\vec p - \vec p')$. Therefore the marginal probability distribution for $\vec p$ is given by:
\begin{equation}
{\cal P}_1 \left(\vec p',t \right) = \int d \vec x \int d \vec p \hspace{0.2 cm} F^B ( \vec x, \vec p ,t) \delta_*^B (\vec p - \vec p',t).
\end{equation}
This expression coincides with ${\cal P}_1 (\vec p',t)$ in eq.(15), but not with ${\cal P}_2 (\vec p',t)$. For a generic operator $\hat A$ and an eigenstate $|a>$, we get from (19,34):
\begin{equation}
\delta_*^B (\tilde A_B -a,t) \equiv V_B \left( |a><a| \right) = f^{-1}_B \left( - i \vec{\nabla}_x , - i \vec{\nabla}_p ,t \right) \delta^W_* (\tilde A_W-a),
\end{equation}
from which we can compute the probability distribution for $\hat A$ by substituting $f=f_B$ in eq.(20).

\subsection{Dynamics and the Heisenberg picture}

From eqs.(32,34) one realizes that, in the Schr\"odinger picture, where the observables remain unaltered and the wave function evolves in time, the Bohm-symbol will also evolve. This is clear from (30,31) as well. We conclude that, contrary to what happens in the Wigner-Weyl case, the time-evolution of the distribution function (eq.(27)) is not sufficient to pin down the complete dynamics. On the other hand, the Heisenberg picture \cite{Dias2} seems to be more adjusted to the de Broglie-Bohm case. If the wave function does not evolve, then $F^B$ and $f_B$ do not evolve either and the complete dynamics is encapsulated in the time evolution of the Bohm-symbols, according to \cite{Lee1, Cohen}:
\begin{equation}
\begin{array}{c}
\frac{\partial \tilde A_B}{\partial t} (\vec x, \vec p;t) = \frac{1}{i \hbar} f_B^{-1} \left(- i \vec{\nabla}_x, - i \vec{\nabla}_p \right) \left[ \tilde A_W (\vec x, \vec p;t) , \tilde H_W (\vec x, \vec p) \right]_M= \\
\\
= \frac{2}{\hbar} f_B \left(- i \vec{\nabla}_x^A, - i \vec{\nabla}_p^A \right) f_B \left(-  i \vec{\nabla}_x^H,  - i \vec{\nabla}_p^H \right) f_B^{-1} \left(- i \vec{\nabla}_x^A - i \vec{\nabla}_x^H, - i \vec{\nabla}_p^A - i \vec{\nabla}_p^H \right) \times \\
\\
\times \sin \left[ \frac{\hbar}{2} \left( \vec{\nabla}_x^A \vec{\nabla}_p^H - \vec{\nabla}_p^A \vec{\nabla}_x^H \right) \right] \tilde A_B \tilde H_B = \left[ \tilde A_B (\vec x , \vec p ,t), \tilde H_B (\vec x , \vec p)\right]_B. 
\end{array}
\end{equation}

\section{Example}

The previous results will be illustrated  with the example of a coherent state. We shall adopt the Heisenberg picture, where the wavefunction is time independent:
\begin{equation}
\psi (x) = \frac{1}{\left[ 2 \pi ( \Delta x)^2 \right]^{\frac{1}{4}}} \exp \left[ - \frac{(x- x_0)^2}{4 ( \Delta x)^2} + \frac{i}{\hbar} p_0 x \right].
\end{equation}
Since $S(x) = p_0 x$, a trivial calculation, yields (cf.(26,32)):
\begin{equation}
f_B (\xi , \eta) = f_B (\eta) = \exp \left[ \frac{\hbar^2 \eta^2}{8 ( \Delta x)^2 } \right],
\end{equation}
and:
\begin{equation}
F^B (x,p) =  \frac{1}{\left[ 2 \pi ( \Delta x)^2 \right]^{\frac{1}{2}}} \delta (p - p_0) \exp \left[ - \frac{(x- x_0)^2}{2 ( \Delta x)^2}  \right].
\end{equation}
From (41) it is easy to obtain:
\begin{equation}
{\cal P}_1 (x) = \frac{1}{\left[ 2 \pi ( \Delta x)^2 \right]^{\frac{1}{2}}} \exp \left[ - \frac{(x- x_0)^2}{2 ( \Delta x)^2} \right] = R^2 (x) , \qquad {\cal P}_2 (p) = \delta (p - p_0).
\end{equation}
Let us now compute $\delta_*^B (p - p')$ from eq.(42). Notice that $f^{-1}_B (\eta) = e^{- \frac{\hbar^2 \eta^2}{8 (\Delta x)^2}}$.
\begin{equation}
\begin{array}{c}
\delta_*^B (p - p') = e^{\frac{\hbar^2}{8 ( \Delta x)^2} \frac{\partial^2}{\partial p^2}} \delta(p-p')= \frac{1}{2 \pi} \int dk \hspace{0.3 cm} e^{\frac{\hbar^2}{8 ( \Delta x)^2} \frac{\partial^2}{\partial p^2}} \hspace{0.3 cm} e^{ik (p-p')} = \\
\\
= \frac{1}{2 \pi} \int dk \hspace{0.3 cm} e^{- \frac{\hbar^2 k^2}{8 ( \Delta x)^2} + ik (p-p')} = \left[ \frac{2 ( \Delta x)^2}{ \pi \hbar^2 } \right]^{\frac{1}{2}} \exp \left[ - 2 \frac{(\Delta x)^2}{\hbar^2} (p - p')^2 \right].
\end{array}
\end{equation}
Let us verify that this is indeed a solution of the star-genvalue equation (cf.(18)):
\begin{equation}
\begin{array}{c}
p *_B \delta_*^B (p-p') =  p \delta_*^B (p - p') - i \frac{\partial f_B^{-1}}{\partial \eta} \left( 0 , - i \frac{\partial}{\partial p} \right) f_B \left( 0 , - i \frac{\partial}{\partial p} \right) \delta_*^B (p -p')=\\
\\
= p \delta_*^B (p -p') + \frac{\hbar^2}{4 ( \Delta x)^2} \frac{\partial}{\partial p} \delta_*^B (p - p')=\\
\\
= p \delta_*^B (p-p') +  \frac{\hbar^2}{4 ( \Delta x)^2} \left[ - 4 \frac{ ( \Delta x)^2}{\hbar^2} ( p - p') \right] \delta_*^B (p -p') = p' \delta_*^B (p -p').
\end{array}
\end{equation}
And so, $\delta_*^B (p-p')$ is indeed the solution of the stargenvalue equation, as expected. Substituting (50) in eq.(43), yields:
\begin{equation}
{\cal P}_1 (p') = \int dx \int dp \hspace{0.3 cm} F^B (x,p) \delta_*^B (p- p') = \left[\frac{2 (\Delta x)^2}{ \pi \hbar^2} \right]^{\frac{1}{2}} \exp \left[- 2 \frac{(\Delta x)^2}{\hbar^2} (p'- p_0)^2 \right],
\end{equation}
which coincides exactly with $| \phi (p')|^2$, where $\phi (p')$ is the Fourier transform of (46). Let us now compute the time evolution (45) of a generic observable $\tilde A_B (x,p)$ induced by a Hamiltonian with Weyl symbol $\tilde H_W (x,p) = \frac{p^2}{2m} + V(x)$:
\begin{equation}
\frac{\partial }{\partial t} \tilde A_B (x,p;t) = \frac{1}{i \hbar} \exp \left(\frac{\hbar^2}{8 ( \Delta x)^2} \frac{\partial^2}{ \partial p^2} \right) \hspace{0.3 cm} \left[\tilde A_W (x,p;t) , \frac{p^2}{2m} + V(x) \right]_M.
\end{equation}
First of all notice that:
\begin{equation}
\begin{array}{c}
\frac{1}{i \hbar} f_B^{-1} \left( - i \frac{\partial}{\partial x} , - i \frac{\partial}{\partial p} \right) \hspace{0.3 cm} \left[\tilde A_W (x,p;t) , \frac{p^2}{2m} \right]_M = \frac{1}{m} f_B^{-1} \left( - i \frac{\partial}{\partial x} , - i \frac{\partial}{\partial p} \right) \hspace{0.3 cm} \frac{\partial \tilde A_W}{\partial x} p= \\
\\
=  \frac{p}{m} f_B^{-1} \left( - i \frac{\partial }{\partial x} , - i \frac{\partial }{\partial p} \right) \frac{\partial \tilde A_W}{\partial x} -  \frac{i}{m} \frac{\partial  f_B^{-1}}{\partial \eta} \left( - i \frac{\partial }{\partial x} , - i \frac{\partial }{\partial p} \right) \frac{\partial \tilde A_W}{\partial x} = \frac{p}{m} \frac{\partial \tilde A_B}{\partial x} + \frac{\hbar^2}{4 m( \Delta x)^2} \frac{\partial^2 \tilde A_B}{\partial x \partial p}.
\end{array}
\end{equation}
Similarly:
\begin{equation}
\frac{1}{i \hbar} f_B^{-1} \left(- i \frac{\partial }{\partial x} , - i \frac{\partial }{\partial p} \right) \hspace{0.3 cm} \left[ \tilde A_W (x,p;t) , V(x) \right]_M = \frac{1}{i \hbar} \left[ \tilde A_B (x,p;t) , V(x) \right]_M,
\end{equation}
where we used the fact that $f_B$ is independent of $\xi$.  
Consequently:
\begin{equation}
\frac{\partial }{\partial t} < \hat A(t) > = \int dx \int d p \hspace{0.3 cm} R^2 (x) \delta (p- p_0)  \left( \frac{\partial \tilde A_B}{\partial p} \frac{p}{m} + \frac{\hbar^2}{4 m( \Delta x)^2} \frac{\partial^2 \tilde A_B}{\partial x \partial p} + \frac{1}{i \hbar} \left[ \tilde A_B  , V(x) \right]_M \right).
\end{equation}
Since $\frac{\partial R^2 }{\partial x} = 2 R \frac{\partial R}{\partial x} = R^2 (x) \frac{(x-x_0)}{(\Delta x)^2}$, we get after an integration by parts:
\begin{equation}
\frac{\partial }{\partial t} < \hat A(t) > = \int dx \int d p \hspace{0.3 cm} R^2 (x) \delta (p- p_0)  \left(  \frac{\hbar^2 (x-x_0)}{4 m( \Delta x)^4}  \frac{\partial \tilde A_B}{\partial p} + \frac{1}{i \hbar} \left[ \tilde A_B  , \frac{p^2}{2m} + V(x) \right]_M \right).
\end{equation}
Now notice that: $- \frac{\hbar^2 (x-x_0)}{4 m( \Delta x)^4} = \frac{\partial}{\partial x} \left[ \frac{\hbar^2}{4 m( \Delta x)^2} - \frac{\hbar^2 (x-x_0)^2}{8 m( \Delta x)^4} \right] = \frac{\partial Q}{\partial x} (x)$, where $Q(x)$ is the quantum potential associated with (46). Finally, we get:
\begin{equation}
i \hbar \frac{\partial }{\partial t} < \hat A(t) > = \int dx \int d p \hspace{0.3 cm} F^B (x,p)     \left[ \tilde A_B (x,p;t)  , \frac{p^2}{2m} + V(x) + Q(x) \right]_M .
\end{equation}
Notice that this formula is valid both for the observable $\hat A$ as well as for the stargenfunction associated with the projector $|a><a|$. In the latter case $\tilde A_B$ would stand for the $*$-genfunction $\delta^B_* (\tilde A_B-a)$, and (58) would yield the time evolution of the probability functional, i.e. ${\cal P}_1 \left( A(t)-a \right) = < \delta^B_* \left( \tilde A_B (t) -a \right)>$. We have thus successfully constructed a Heisenberg picture for the system. This can be achieved for any system. However, it is important to emphasize that the simple formula (58) only applies for $f_B$ of the form (47). For more general cases, equation (58) becomes more intricate, although formally one just has to replace the Moyal bracket by the de Broglie-Bohm bracket (cf.(45)) in equation (58) to obtain the time evolution. 
    
\section{Conclusions}

We have established a connection between the de Broglie-Bohm interpretation and the quasi-distribution formulation of quantum mechanics. For the theorist familiar with the latter formulation, the causal interpretation corresponds (at least formally) to just another quasi-distribution. Consequently, the whole machinery for calculating average values, probability functionals and the dynamics can be trivially applied. 

Our aim was not to construct a proper causal interpretation in non-commutative quantum phase space. That has been the purpose of some attempts in the past, \cite{Lee2}-\cite{Polavieja}. However, to our result that the de Brolie-Bohm interpretation can be implemented via a quasi-distribution there is a converse result. Namely, that any quasi-distribution can be cast in a form almost identical to the causal form (26), except that the underlying variables $\vec x$ and $\vec p$ no longer commute. The Wigner function can then be regarded as a $\hbar$-deformation of the Bohm function (26). This will be the subject of a future work \cite{Dias3}.

\vspace{1 cm}

\begin{center}

{\large{{\bf Acknowledgments}}} 

\end{center}

\vspace{0.3 cm}
\noindent
We would like to thank Jo\~ao Marto for useful suggestions and for reading the manuscript. This work was partially supported by the grants ESO/PRO/1258/98 and CERN/P/Fis/15190/1999.

\end{document}